\documentclass[english,prd,superscriptaddress,nofootinbib,preprintnumbers,twocolumn,showpacs]{revtex4}
\usepackage[utf8]{inputenc}
\usepackage[english]{babel}
\usepackage{amsmath}
\usepackage{amsfonts}
\usepackage{amssymb}
\usepackage{epsfig}
\usepackage{graphics,psfrag,rotating}
\usepackage{graphicx}
\usepackage{dcolumn}
\usepackage{bm}
\usepackage{epstopdf}
\usepackage{color}
\usepackage[usenames,dvipsnames,svgnames]{xcolor}
\usepackage[T1]{fontenc}
\usepackage{multirow}
\usepackage{float}

\usepackage{subfigure}

\usepackage{enumitem}
\usepackage[colorlinks=true,
            linkcolor=red,
          urlcolor=gray,
            citecolor=blue]{hyperref}

\def\3nab{\tilde{\nabla}}

\def\be {\begin{equation}}
\def\ee {\end{equation}}
\def\ba {\begin{align}}
\def\ea {\end{align}}

\def\bc {\begin{center}}
\def\ec {\end{center}}
\def\case#1/#2{\frac{#1}{#2}}

\newcommand{\bver}{\begin{verbatim}}
\newcommand{\ever}{\end{verbatim}}

\newcommand{\bea}{\begin{eqnarray}}
\newcommand{\eea}{\end{eqnarray}}
\newcommand{\beaa}{\begin{eqnarray*}}
\newcommand{\eeaa}{\end{eqnarray*}}

\def\case#1/#2{\textstyle\frac{#1}{#2}}

\begin{document}
%%%%%%%%%%%%%%%%%%%%%%%%%%%%%%%%%%%%%%%

\title{Theoretical and observational constraints of viable $f(R)$ theories of gravity}

\author{\'Alvaro de la Cruz-Dombriz\footnote{ alvaro.delacruzdombriz [at] uct.ac.za }}
\affiliation{Astrophysics, Cosmology and Gravity Centre (ACGC), Department of Mathematics and Applied Mathematics, University of Cape Town, Rondebosch 7701, Cape Town, South Africa.}
\author{Peter K. S. Dunsby\footnote{ peter.dunsby [at] uct.ac.za}}
\affiliation{Astrophysics, Cosmology and Gravity Centre (ACGC), Department of Mathematics and Applied Mathematics, University of Cape Town, Rondebosch 7701, Cape Town, South Africa.}
\affiliation{South African Astronomical Observatory,  Observatory 7925, Cape Town, South Africa}
\author{Sulona Kandhai\footnote{kndsul001[at] myuct.ac.za}}
\affiliation{Astrophysics, Cosmology and Gravity Centre (ACGC), Department of Mathematics and Applied Mathematics, University of Cape Town, Rondebosch 7701, Cape Town, South Africa.}
\author{Diego S\'aez-G\'omez\footnote{ dsgomez [at] fc.ul.pt}}
\affiliation{Departamento de F\'isica \& Instituto de Astrof\'isica e Ci\^encias do Espa\c{c}o,
Faculdade de Ci\^encias da Universidade de Lisboa, Edif\'icio C8, Campo Grande, P-1749-016
Lisbon, Portugal}

\pacs{04.50.Kd, 98.80.-k, 98.80.Cq, 12.60.-i}

\begin{abstract} 
Modified gravity has attracted much attention over the last few years and remains a potential candidate for dark energy. In particular, the so-called {\it viable} $f(R)$ gravity theories, which are able to both recover General Relativity (GR) and produce late-time cosmic acceleration, have been widely studied in recent literature. Nevertheless, extended theories of gravity suffer from several shortcomings which compromise their ability 
to provide realistic alternatives to the standard cosmological $\Lambda$CDM Concordance model.
We address the existence of cosmological singularities and the conditions that  guarantee late-time acceleration,
assuming reasonable energy conditions for standard matter in the so-called Hu-Sawicki $f(R)$ model, currently among the most widely studied modifications to General Relativity. 
Then using the Supernovae Ia Union 2.1 catalogue, we further constrain the free parameters of this model. The combined analysis of both theoretical and observational constraints sheds some light on the viable parameter space of these models and the form of the underlying effective theory of gravity. 
\end{abstract} 
%%%%%%%%%%%%%%%%%%%%%%%%%%%%%%%%%%%%%
%\date{\today}
\maketitle

%%%%%%%%%%%%%%%%%%%%%%%%%%%%%%%%%%%%%
\section{Introduction}
%%%%%%%%%%%%%%%%%%%%%%%%%%%%%%%%%%%%%

Over the last few years, a great deal of effort related to the problem of the origin of late-time cosmic acceleration has been devoted to the so-called $f(R)$ theories of gravity. This is due to the fact that by a choosing the Lagrangian of the gravitational interaction to be an appropriate function of the Ricci scalar, the late-time acceleration of the universe expansion can be reproduced without the need of introducing a dark energy field (for a review see Ref.~\cite{Nojiri:2010wj}). Contrary to the first $f(R)$ models from the 80's, for example  Starobinsky's $R^2$ inflation model, much of the current work on $f(R)$ gravity is aimed at obtaining a description of the late-time history of the universe, when the curvature is very small. Roughly speaking, these models provide an infrared correction to General Relativity (GR), which may be inspired by string theories %Ref.~
\cite{Nojiri:2006je}. Moreover, the analysis of inflation within the framework of modified gravity and even the unification of late-time acceleration and inflation still draws a great deal of attention, particularly after the last release of Planck data and the success of Starobinsky inflation \cite{Planck-Inflation}. It is therefore possible that these types of modifications to general relativity could lead to a complete picture of the evolution of the universe %Ref.~
\cite{staro}. Unfortunately, in general, such modifications of GR are plagued by a number of problems, such as violations of local gravity tests, the absence of a matter dominated era and antigravity regimes among others. In order to deal with these shortcomings, some $f(R)$ models referred to as  {\it viable} have been proposed in the last few years (see Refs.~\cite{Hu:2007nk,Starobinsky:2007hu}). Those models are able to introduce corrections at cosmological scales, while GR is recovered on local scales and the usual predictions of GR remain the same. To do so, the authors of these works extended the so-called Chameleon mechanism \cite{Khoury:2003rn}, initially applied to scalar-tensor theories, to $f(R)$ gravity. The Chameleon mechanism basically introduces a scale hierarchy over the additional terms of the gravitational action so that on local scales, for example the Earth or the Solar System, GR is effectively recovered. On the other hand, these terms become important on cosmological scales, whereby the appropriate choice of theory parameters, the late-time acceleration can be reproduced. In addition, these viable modified theories of gravity are also able to evade the Ostrogradski and Dolgov-Kawasaki instabilities \cite{Dolgov:2003px}. All these features make these theories a promising candidate for dark energy. 

However, a common issue of every viable $f(R)$ gravity is the presence of a number of theoretical shortcomings such as back-reaction averaging effects, absence of smooth junction conditions in astrophysical context \cite{CDGN,CD}, faster growth of structures in disagreement with large-scale structure catalogues \cite{Dombriz-perturbations}, the appearance of unexpected singularities and existence of anti-gravity regimes, among others. The latter two issues form part of the study presented in this manuscript. 
In more detail, a common feature of every viable $f(R)$ gravity is the presence of sudden cosmological singularities both in the past and the future (for a classification of singularities, see Ref.~\cite{Nojiri:2005sx}). The existence of such singularities within General Relativity is connected to violations of the energy conditions by the matter content, particularly dark energy. Nevertheless, within modified gravity, the energy conditions may be violated naturally through the extra geometrical terms that appear in the field equations \cite{Nojiri:2008fk}. The occurrence of such singularities have been explored extensively in the literature, since observations do not discard an equation of state parameter for dark energy $w_{}<-1$ \cite{SingOc}. One of these cosmological singularities is the so-called {\it sudden} singularity, where the first derivative of the Hubble parameter diverges \cite{Barrow:2004xh}. Viable $f(R)$ gravity, in general contains this type of cosmological singularity. Furthermore, the singularity represents an asymptotically stable point and therefore its avoidance depends entirely upon the initial conditions and the election of the free parameters (see Refs.~\cite{Appleby:2009uf,SaezGomez:2012ek}). 
%Alternative ways of solving such issue propose slight modifications of the original action which keep its main properties, particularly adding a quadratic term in the action \cite{Appleby:2009uf}. {\color{red} Maybe a bit more on singularities?}.

%Concerning the attractiveness of gravitational interaction in the $f(R)$ theories context
On the other hand, it is well known that in the context of GR without a cosmological constant, a non-positive contribution for the space-time geometry to the Raychaudhuri equation, or in other words the attractiveness of gravitational interaction, is obtained once standard fluids are assumed and regardless of the solution of the Einstein's equations \cite{HE, Wald}. However, this result can be reversed in the context of extended theories of gravity, where depending on the theory and parameter choice, the subsequent convergence (or divergence) of geodesics for fundamental observers can be obtained without invoking the presence of exotic fluids. 
Moreover, an upper bound to the contribution of space-time geometry can be provided both in terms of the gravitational model and the metric under consideration. Using this upper bound and assuming usual energy conditions, restrictions on $f(R)$ models can be derived in order to constrain their cosmological viability \cite{Dombriz_JCAP_Attractive}.
Consequently, the careful analysis of the geometrical terms in the Raychaudhuri equation for extended theories of gravity plays a critical role in the demonstration of the singularities theorems \cite{HE}, as well as in the context of the so-called Holographic Principle \cite{Bousso}. 
%
%Very often it is claimed that these theorems require energy conditions to hold, since for instance the SEC as we have just seen implies $R_{\mu\nu}\xi^a\xi^b\geq0$. However, these theorems are essentially mathematical theorems independent of the gravitational theory. Energy conditions are necessary for these theorems to hold if and only if GR is assumed. 
%
An analysis of this geometrical contribution \cite{ACD} showed that it can be interpreted as the mean curvature in the direction of the congruence \cite{Eisenhart}. It can also be easily verified that for a Robertson-Walker model with a negative deceleration parameter, this contribution is positive \cite{ACD} and the attractive character of gravity vanishes. % From this analysis, it is clear that the accelerated expansion of the Universe may be in conflict with
%
%The energy conditions for different modified gravity theories have been widely studied in the literature \cite{Santos, Others_Energy_Conditions}, where the most general approach consists of defining an effective stress-energy tensor by analogy with General Relativity, including all extra geometrical terms in order to obtain an expression for the mean curvature. %Then, generalized energy conditions on this non-physical effective tensor were imposed.  
%
%This generalization of the energy conditions {\color{red} remains doubtful} \cite{Santos, ACD} since there is no natural motivation other that making a direct analogy with General Relativity. %This extension is in fact only motivated when the new terms appearing in the field equations are identified with physical fields. Nevertheless, these new terms may be understood as possessing only a geometrical meaning. Thus, there is no reason to assume any energy conditions on these terms. 
%
Actually. as shown in \cite{ACD}, the mean curvature for a given geodesics turns out to be positive for almost all timelike directions in a Robertson-Walker model with the present value of the deceleration parameter. 
%Therefore, if these extended energy conditions hold, the present accelerated expansion of the Universe cannot be accommodated.
%
%The analysis in this paper precisely shall focus on not assuming any conditions for non-physical fluids and studying the upper bound for the extra geometrical contributions to the Raychaudhuri equation.

The present manuscript is devoted to the analysis of a class of viable $f(R)$ models -- the so-called Hu-Sawicki gravity model \cite{Hu:2007nk} -- which has received a lot of attention lately \cite{Tsujikawa:2007xu}. Here we investigate the possible constraints on the free parameters of this model by using both theoretical limits and observational data, particularly the Union2.1 catalogue of Supernovae Ia \cite{Suzuki:2011hu}. Then, the free parameters are constrained by obtaining the region of parameter-space which is free of cosmic singularities and also has a positive contribution to the Raychaudhuri equation at late times. Initial conditions for the background evolution are fixed at large ($z\sim 10$) redshifts to be the same as in the $\Lambda$CDM model in order to guarantee that the high redshift cosmology is compatible with BBN and CMB constraints. After obtaining a region free of singularities, which provides a smooth evolution from the past until today, the free parameters are then fitted by using Supernovae Ia data. 
%Also the antigravity regions of the parameter space are analysed... {\color{red}(To be continued...)}.\\

The paper is organised as follows: In Sec. \ref{Section_General} we provide an overview of $f(R)$ theories of gravity in the metric formalism in general and the Hu-Sawicki model in particular, providing a brief review of the dynamical system approach which enables us to easily solve the background cosmological equations. Then in Sec. \ref{Section_Singularities} we discuss the emergence of sudden singularities in these kind of models using the equivalent picture of scalar-tensor theories. We also present the theoretical analysis leading to upper bounds on the positive geometrical contributions to the Raychaudhuri equation for $f(R)$ models. The statistical analysis using supernovae data is performed in Sec. \ref{Section_SN} enabling us, together with other gravitational and cosmological tests to constrain the viable parameters space.  We end the paper in Sec. \ref{Section_Conclusions} presenting the main results of this investigation. A brief appendix \ref{Appendix_I} at the end of the paper gives details on the process to find the apparent magnitude statistical minimum.
Unless otherwise specified,  natural units ($\hbar=c=k_{B}=8\pi G=1$) will be used throughout this paper.

%%%%%%%%%%%%%%%%%%%%%%%%%%%%%%%%%%%%%
\section{Cosmological evolution in Hu-Sawicki $f(R)$ model}
\label{Section_General}
%%%%%%%%%%%%%%%%%%%%%%%%%%%%%%%%%%%%%
$f(R)$ gravity usually refers to a set of theories whose Lagrangian is given by a general function of the Ricci scalar,
\be
S=\int {\rm d}^4x \sqrt{-g}\left[f(R)+2\mathcal{L}_m\right] \ ,
\label{1.1}
\ee
where %$\kappa^2$ is the gravitational coupling constant and 
$\mathcal{L}_m$ is the Lagrangian of the matter content. It is straightforward to obtain the field equations  by varying the action with respect to the metric field $g_{\mu\nu}$, leading to 
\be
R_{\mu\nu} f_R- \frac{1}{2} g_{\mu\nu} f(R) + \left(g_{\mu\nu}  \Box -  \nabla_{\mu} \nabla_{\nu}\right)f_R\,=\,T^{(m)}_{\mu\nu}\ .
\label{1.2}
\ee
where $f_R\equiv \frac{{\rm d}f}{{\rm d}R}$.  Higher derivatives of $f$ with respect to $R$ will be denoted as $f_{2R}$, $f_{3R}$, etc.

%As is covered by previous literature, cosmic acceleration can indeed be achieved by the appropriate choice of the $f(R)$ Lagrangian. 

We are primarily interested in studying spatially flat Robertson-Walker cosmologies, whose metric, expressed in the usual co-moving coordinates, is given by
\be
{\rm d}s^2=-{\rm d}t^2+a(t)^2\sum_{i=1}^3\left({\rm d}x^{i}\right)^2\ .
\label{1.4}
\ee
Then, the corresponding field equations obtained from (\ref{1.2}) and corresponding to a dust-dominated Universe become   
\bea
H^2&=&\frac{1}{3f_R}\left( \rho_m +\frac{Rf_R-f}{2}-3H\dot{R}f_{2R}\right), \label{Friedmann} \\
-3H^2-2\dot{H}&=&\frac{1}{f_R}\left[%p_m+
\dot{R}^2f_{3R}+\left(2H\dot{R}+\ddot{R}\right)f_{2R}\right.\nonumber\\
&+&\left.\frac{1}{2}(f-Rf_R)\right]\,,
\label{Raychaudhuri}
\label{1.5}
\eea
where the Hubble parameter is $H(t)=\dot{a}/a$, the dot denotes a derivative with respect to cosmic time, and $\rho_m$ %and $p_m$ 
denotes the standard matter energy density. % and pressure respectively. 
% and the Ricci scalar is given by $R=6\ (2H^2+\dot{H})$. 
%
We can also use the continuity equation,
\bea \dot{\rho}_m+3H %(1+w_m)
\rho_m=0 \label{continuity} \eea 
to reduce the number of independent equations.

It has been shown recently \cite{DS} that it is convenient (and numerically more stable) to express the cosmological equations as a set of autonomous first order equations in order to study the expansion history of a general class of $f(R)$ theories. Taking advantage of this fact, we rewrite equations (\ref{Friedmann}) -  (\ref{continuity}) in terms of the following dynamical system variables
\begin{align}\label{ncvariables}
{x}&= \frac{\dot{R}f_{2R}}{H f_{R}},~~~~{v}=\frac{R}{6H^{2}}, \nonumber\\
\\
{y}&=\frac{f}{6H^2 f_{R}}, ~~~~{\Omega} =\frac{\rho_{m}}{3H^2 f_{R}}.      \nonumber    
\end{align}
Written in terms of (\ref{ncvariables}), the Friedmann and Raychaudhuri equations become 
\bea 
1={\Omega} + {v} -{x} -{y} \\
\nonumber\\
\frac{{\rm d}h}{{\rm d}z} = \frac{h}{z+1} \left( 2- {v} \right)\;,\label{dhdz}
\eea
where $h=H/H_{0}$ and we obtain the following  set of first order differential equations directly from the dynamical system variables.
\begin{align}
\frac{{\rm d} {x}}{{\rm d}z}=&\frac{1}{(z+1)}\left[  -  {\Omega}+{ {x}}^{2}+ \left( 1+ {v} \right)  {x} -2 {v}+4 {y}\right] , \label{Neqx}\\
& \nonumber \\
\frac{{\rm d} {y}}{{\rm d}z}=&-\frac{1}{(z+1)}\left({ {v} {x}{\Gamma}- {x} {y}+4\, {y}-2\, {y} {v}}\right)\;,\\
& \nonumber \\
\frac{{\rm d} {v}}{{\rm d}z}=&-\frac{ {v}}{(z+1)} { \left(  {x}{\Gamma}+4-2\, {v} \right) } ,\label{Neqv}\\
& \nonumber \\
\frac{{\rm d} {\Omega}}{{\rm d}z}=&\frac{1}{(z+1)}\left[{ {\Omega}\, \left( -1+ {x}+2\, {v} \right) }\right]\;.
\label{dodz} 
\end{align}
These equations describe the cosmological evolution of a general $f(R)$ theory of gravity, where $\Gamma \equiv \frac{f_{R}}{Rf_{2R}}$ specifies the theory \cite{DS_Hu-Sawicki}. 

In this paper we focus on the analysis of the so-called viable $f(R)$ theories of gravity, which, in addition to producing the late time accelerated era of expansion, also recovers results consistent with General Relativity on local scales. To illustrate our analysis, let us consider the model of this kind proposed in Ref.~\cite{Hu:2007nk},
\be
f(R)=R-cH_0^2\frac{b(R/cH_0^2)^n}{d(R/cH_0^2)^n+1}\equiv R+f^
{HS}(R)\ ,
\label{1.3}
\ee
where $\{b,c,d,n\}$ are constants to be determined by both theoretical and observational constraints, while $H_0$ is the $\Lambda$CDM Hubble parameter evaluated today. For this model, the $\Gamma$ term in the dynamical systems equations takes the following form
\bea 
\Gamma = -\frac{\left( d r^n + 1 \right) \left[ r\left( d r^{n} +1\right)^{2} - bnc r^{n} \right] }{bnc\left[ r^{n}(n-1) - d r^{2n}(n+1)  \right]}\;,
\eea
where $r=R/cH_0^2$ is the dimensionless Ricci scalar. % and $R_0$ is the present day $\Lambda$CDM Ricci scalar. 

The success of this model lies in its ability to produce an effective cosmological constant at late times, thus mimicking the expansion history of the $\Lambda$CDM model, as well as avoiding violations of local gravity tests. To do so, the extra scalar degree of freedom - known as the scalaron - behaves like a Chameleon field, whose mass is given by:
\be
m_{f_{}^{HS}}=\sqrt{3f_{2R}^{HS}}\,.
\label{1.3a}
\ee
Summarising, since the mass of the scalaron (\ref{1.3a}) depends on the scale via the Ricci scalar, roughly speaking, the so-called thin-shell condition (a smooth transition from high to low curvature regimes) is satisfied provided the mass (\ref{1.3a}) is large enough in the high curvature regime, such that deviations from General Relativity are avoided. For further details about the chameleon mechanism {\it c.f.} Ref.\cite{Khoury:2003rn} and for its extension to $f(R)$ gravity, see Ref.~\cite{Hu:2007nk}. 

In spite of the great success of models of this kind, they are plagued with several shortcomings such as the presence of antigravity regimes or the occurrence of cosmic curvature singularities. Both of these issues are analysed in detail below, but before doing so, let us first illustrate their cosmological behaviour, given by (\ref{1.3}) and how they are able to mimic the cosmological constant behaviour at late times. 

\begin{figure}
\begin{center}
\includegraphics[width=0.4\textwidth]{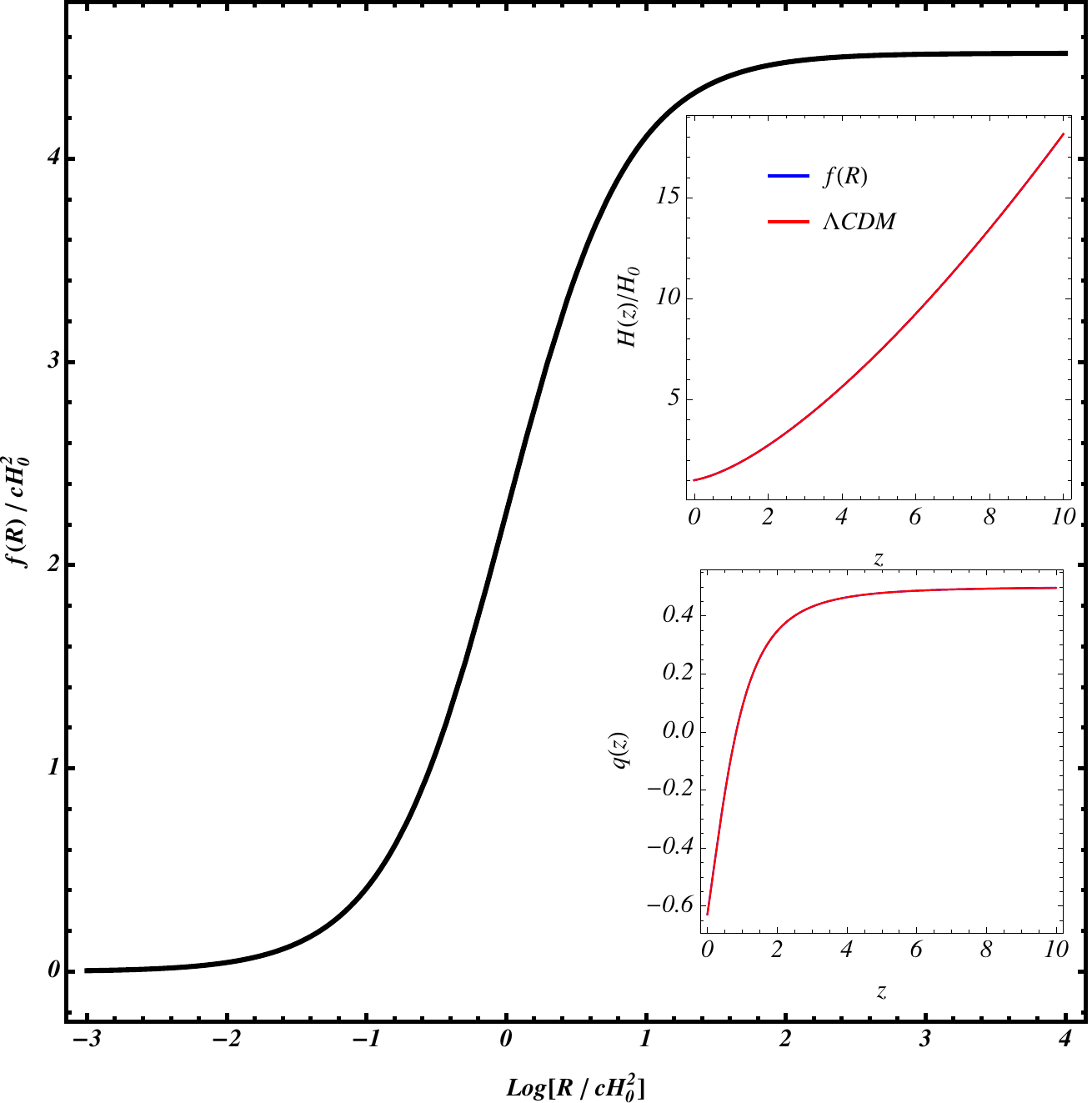}
\end{center}
\caption{The Hu-Sawicki model for a sample of the free parameters $\{n=1,b=200\}$. We show the redshift evolution of the Hubble parameter and the deceleration parameter in the inner upper panel and inner lower panel, respectively, for the Hu-Sawicki (blue) model and the $\Lambda$CDM model (red). Both the Hubble parameter and the deceleration parameter corresponding to this model are clearly indistinguishable from $\Lambda$CDM.}
\label{fig1}
\end{figure}

In order to illustrate this qualitatively, Fig. \ref{fig1} depicts the shape of $f^{HS}(R)$ for a set of the free parameters of the model.  The free parameter $n$ controls the slope of the transition to a constant {\it plateau}. The amplitude of the correction is directly determined by the free parameter $c$, so that when $R\ll cH_0^2$, corrections to GR are negligible, and in the high curvature limit,  $R\gg cH_0^2$, $f^{HS}(R)$ behaves effectively like a cosmological constant, namely\footnote{In expression (\ref{HS_assymptotic}) the limit must be understood as $R\gg cH_0^2$, i.e., eras with high Ricci curvature, such as matter/radiation dominated eras.} 
 \bea
 \lim_{cH_0^2/R \rightarrow 0} f^{HS}(R) \approx  - \frac{b}{d}m^{2}\;.
 \label{HS_assymptotic}
 \eea
As was originally presented in \cite{Hu:2007nk}, we limit the choices of the free parameters by requiring that this theory must mimic the $\Lambda$CDM model. In order for this to occur, we require that 
 \bea
 c=6(1-\Omega_m)\frac{d}{b}\;.
 \eea
In this way, the amplitude of the plateau is controlled by the free parameters $\{b,\, d\}$ and the matter density today $\Omega_m\equiv\rho_{m,0}/3H_0^2$.
 % As shown, the value of the free parameter $c$ controls when the transition occurs, while for $R\ll cH_0^2$ the corrections to GR are negligible, whereas for $R\gg cH_0^2$, $f_{HS}(R)$ behaves essentially as an effective cosmological constant. In the following we are assuming $c=6(1-\Omega_m)d/b$, such that the transition will be controlled by the free parameters $\{b,\, d\}$ and the matter density $\Omega_m=\frac{\rho_0}{3H_0^2}$ {\color{red} (an additional explanation about the election of $c$ should be included here)}. 
 %
 %%% THAT GOES TO THE FIGURE %%%
In the top central panel of Fig.~\ref{fig1}, the Hubble parameter evolution is compared with the $\Lambda$CDM model, while the central bottom panel depicts the deceleration parameter $q=-\ddot{a}/aH^2$. This clearly shows that for this choice of parameters, the expansion history is indistinguishable from the $\Lambda$CDM model, a feature which makes this class of theories such a popular parameterisation of dark energy. Having said this, in what follows, we will show that not all values of the parameters lead to viable expansion histories due to the presence of curvature singularities at physically relevant redshifts nor they guarantee cosmological expansion at late times. 
%As shown, the evolution of the Hubble parameter may not differ specially from the $\Lambda$CDM model, whereas the deceleration parameter may show large deviations. \\

%{\color{red} I would like the X and Y axes characters to be different for $f(R)$ and $log(R)$}

%In the inner upper panel, the Hubble parameter evolution is depicted for the Hu-Sawicki model (blue line) and $\Lambda$CDM model (red line), which are completely indistinguishable. In the inner lower panel, we depict the deceleration parameter $q=-\ddot{a}/H^2a$, both the Hu-Sawicki model (blue) and the $\Lambda$CDM model (red). As above, both models do not show any differences. 

%
%%%%%%%%%%%%%%%%%%%%%%%%%%%%%%%%%%%%%%%%%
\section{Singularities and the non-attractive character of gravity in viable $f(R)$ theories}
\label{Section_Singularities}
%%%%%%%%%%%%%%%%%%%%%%%%%%%%%%%%%%%%%%%%%
%
\subsection{Singularities}
\label{Sub Singularities}
\begin{figure}
\begin{center}
\includegraphics[width=0.4\textwidth]{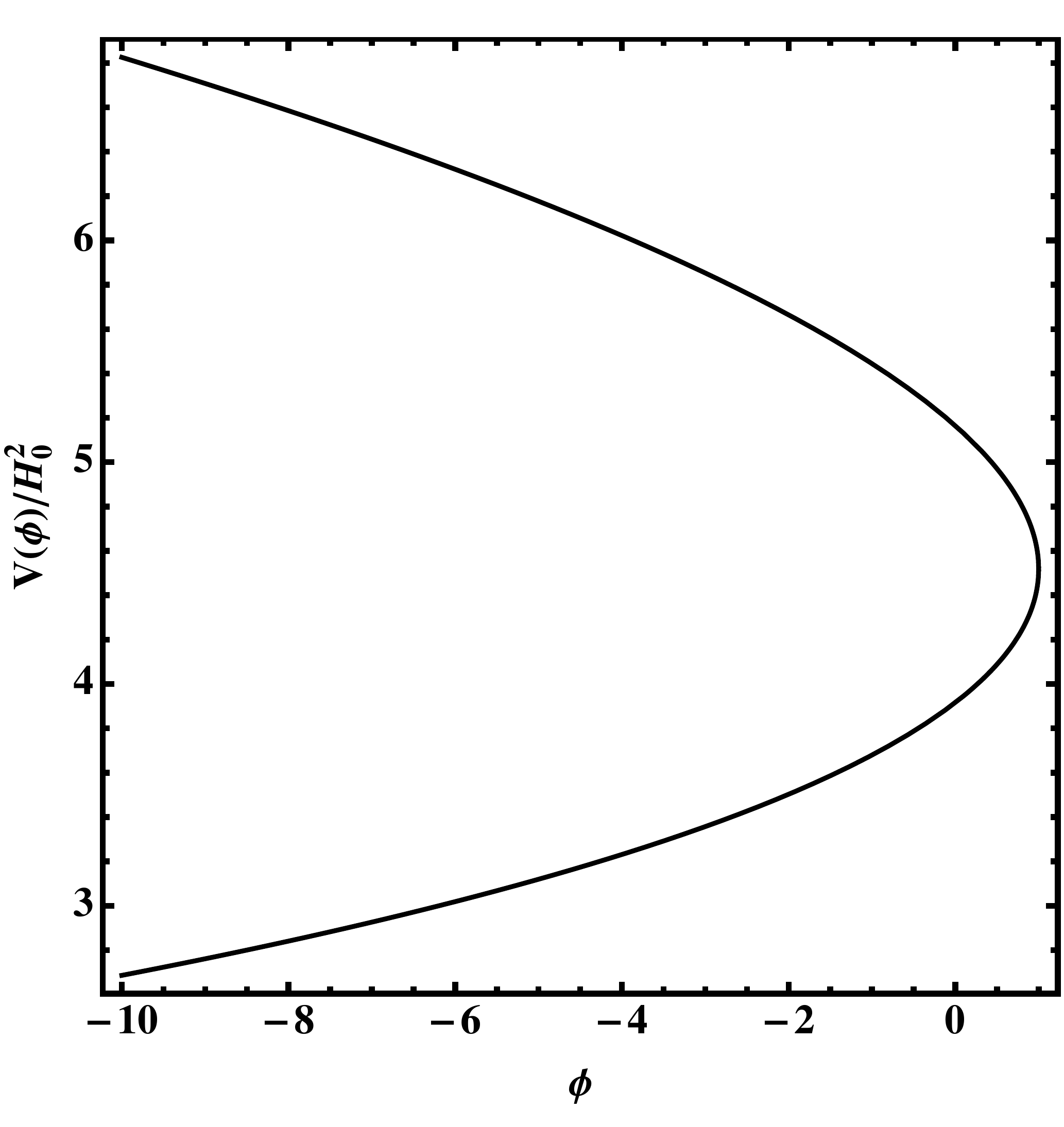}
\end{center}
\caption{Evolution of the scalar potential (\ref{1.11}) for $n=1$. The singular behaviour lies at $\phi=1$. The lower branch is singularity free whereas the upper branch leads inevitably towards the singularity. The potential corresponds to the free parameteres used in Fig.~\ref{fig1}, $\{n=1,b=200\}$.}
\label{fig2}
\end{figure}

%%%%%%%%%%%%%%%%%%%%%%%%%%%%%%%%%%%%%%%%%
One of the main shortcomings of viable $f(R)$ theories of gravity is the occurrence of cosmological singularities, in particularly the appearance of a sudden singularity, where $\dot{H}\rightarrow\infty$ in a finite time $t_s$ (see \cite{Appleby:2009uf,SaezGomez:2012ek}). This is a feature which can be easily analysed within the scalar-tensor framework of $f(R)$ gravity, where the action (\ref{1.1}) takes the form
\be
S=\int {\rm d}^4x \sqrt{-g}\left[\phi R-V(\phi)+2\mathcal{L}_m\right] \ ,
\label{1.8}
\ee
by means of the relations
\be
\phi=f_R\ \quad ; \quad V(\phi)=Rf_R-f(R)\ .
\label{1.9}
\ee
For the model (\ref{1.3}), the scalar field $\phi$ and its potential in terms of the Ricci scalar become
\begin{eqnarray}
\phi&=&1-b n\frac{(R/cH_0^2)^{n-1}}{\left[1+d(R/cH_0^2)^n\right]^2}\ ,\\
 V(\phi(R))&=&\frac{bcH_0^2(R/cH_0^2)^n\left[1-n+d(R/cH_0^2)^n\right]}{\left[1+d(R/cH_0^2)^n\right]^2}\nonumber.
\label{1.10}
\end{eqnarray}
In general it is not possible to get the explicit expression of the scalar potential in terms of the scalar field $V=V(\phi)$ since the first expression in (\ref{1.10}) is not analytically invertible for a general $n$. Nevertheless, this is possible for the case $n=1$, such that the scalar potential yields
\be
V(\phi)=cH_0^2\frac{b+(1-\phi)\pm2\sqrt{b(1-\phi)}}{d}\ .
\label{1.11}
\ee
Note that in this case the potential is not univocally defined, as depicted in Fig.~\ref{fig2}. It is straightforward to check that the sudden singularity, where $R\rightarrow\infty$, occurs for
\be
\phi\rightarrow 1\ , \quad V\rightarrow \frac{bc}{d}H_0^2\ ,
\label{1.12}
\ee
since $\dot{H}\propto V'(\phi)$ and the first derivative of the potential $V'(\phi\rightarrow 1)\rightarrow \infty$.
Hence, in order to construct a consistent and smooth cosmological evolution for the $f(R)$ model (\ref{1.3}), the occurrence of such singularity has to be avoided. 
%In addition, as $\phi=f_R>0$ to avoid antigravity regimes, the following constraint is obtained
%\be
 %0<f_R<a\ \quad \text{or equivalently} \quad -a<f_R^{HS}<0\ ,
 %\label{1.13}
%\ee
%where we have called $f^{HS}(R)=-\frac{b(R/cH_0^2)^n}{d(R/cH_0^2)^n+1}$. %Hence, by calculating the maximum and minimum value of $f_R^{HS}$, we can obtain an additional constraint on the free parameters, which reduces the number of parameters that have to be fit by the observational data.\\
%
%
Note that the two branches of the scalar potential, Fig.~\ref{fig2}, contain different asymptotically stable points. While the upper branch ends at the singular point $\phi=1$, and any cosmological evolution located initially on that branch, the other branch ends in an asymptotically stable de Sitter evolution (see Ref.~\cite{SaezGomez:2012ek}). Therefore, depending upon the initial conditions and the model parameters values, the singularity may be avoided, as shown in the following Section. 

%%%%%%%%%%%%%%%%%%%%%%%%%%%%%%%%%%%%%%%%%%%%%%%%%%%%
\subsection{Attractive character}
\label{Sub Attractive character}
%%%%%%%%%%%%%%%%%%%%%%%%%%%%%%%%%%%%%%%%%%%%%%%%%%%%

In this section we focus on finding inequalities which provide an upper bound for the positive contribution to the space-time geometry of the Raychaudhuri equation for timelike  geodesics\footnote{The analysis for null geodesics is much simpler as shown in \cite{Dombriz_JCAP_Attractive}.}, rendering the gravitational interaction attractive. Let us express the Raychaudhuri equation for timelike geodesics as \cite{Wald,Raychaudhuri}
\begin{eqnarray}
\frac{\text{d}\theta}{\text{d}\tau}=-\frac{1}{3}\,\theta^2-\sigma_{\mu\nu}\sigma^{\mu\nu}+\omega_{\mu\nu}\omega^{\mu\nu}-R_{\mu\nu}\xi^\mu\xi^\nu\,,
\label{ray}
\end{eqnarray}
where $\theta$, $\sigma_{\mu\nu}$ and $\omega_{\mu\nu}$ are respectively the expansion, shear and rotation of the congruence of timelike geodesics generated by the tangent vector field $\xi^\mu$ and $\tau$ is an affine parameter. One of the standard interpretations of the Raychaudhuri equation is that, once the Strong Energy Condition  (SEC) is assumed\footnote{Note that both dust matter and radiation satisfy the SEC. For a discussion about cases where this condition does not hold see \cite{HE}. In particular, a stress-energy tensor corresponding to a cosmological constant $\Lambda$ fluid does not fulfill the SEC.}
\begin{eqnarray}
T_{\mu\nu}\xi^\mu\xi^\nu\geq-\frac{1}{2}T,%\ \ \ \ \ \ \ \ \ \ \ \text{SEC}\,.
\label{SEC}
\end{eqnarray}
Provided that GR is considered as the underlying theory, the SEC immediately implies that $R_{\mu\nu}\xi^\mu\xi^\nu\geq 0$, which may be interpreted as a manifestation of the attractive character of gravity. 
It therefore follows that the mean curvature \cite{ACD, Eisenhart} in every timelike direction defined by
\begin{eqnarray}
{\cal M}_{\xi}\,\equiv\,-R_{\mu\nu}\xi^\mu\xi^\nu
\label{Mean_curvature}
\end{eqnarray}
is negative or zero in GR provided that the SEC holds.
The utility of the Raychaudhuri equation in the singularity theorems is based on the following result: if one chooses a congruence of timelike geodesics  whose tangent vector field is locally hypersurface-orthogonal, then one gets $\omega_{\mu\nu}=0$ for all the congruences. Since the term $\sigma_{\mu\nu}\sigma^{\mu\nu}$ is non-negative and whenever $R_{\mu\nu}\xi^\mu\xi^\nu \geq 0$ is assumed, then
\begin{eqnarray}
\frac{\text{d}\theta}{\text{d}\tau}+\frac{1}{3}\theta^{2}\leq 0\;
%\end{eqnarray}
%which implies
%\begin{eqnarray}
\rightarrow\; \theta^{-1}(\tau) \geq \theta^{-1}_0 +\frac{1}{3}\tau\,.
\end{eqnarray}
This inequality %equation
tells us that a congruence initially converging ($\theta_0\leq 0$) will converge to zero in a finite time. % $\tau \leq 3/\lvert \theta_{0} \rvert$. 
Reverse reasoning backwards in time can be easily formulated. 
%, or in a reversed sense, if the congruence is initially diverging $\theta_{0} \geq 0$ it was focused until zero size in the past. 
%
Let us stress at this stage that the requirement for the previous reasoning to be true for any general theory of gravity does not need any energy condition to hold, but rather that $R_{\mu\nu}\xi^\mu\xi^\nu \geq 0$ for every non-spacelike vector. 
%
%These inequalities provide us with an upper bound for the contribution of space-time geometry to the Raychaudhuri equation for timelike geodesics \eqref{ray}. 
In what follows, we focus on timelike geodesics, referring the reader to \cite{Dombriz_JCAP_Attractive}, where details on null geodesics were presented. We also consider the aforementioned constraint in late-time cosmological scenarios, i.e., assuming a de Sitter phase and subdominant contributions from both radiation and dust. Thus, the Ricci scalar $R=R_0$ will be approximately constant for situations where we require cosmological expansion %attractiveness 
of timelike geodesics in order to match observations.

\begin{figure}[h!]
\begin{center}
\includegraphics[width=0.47\textwidth]{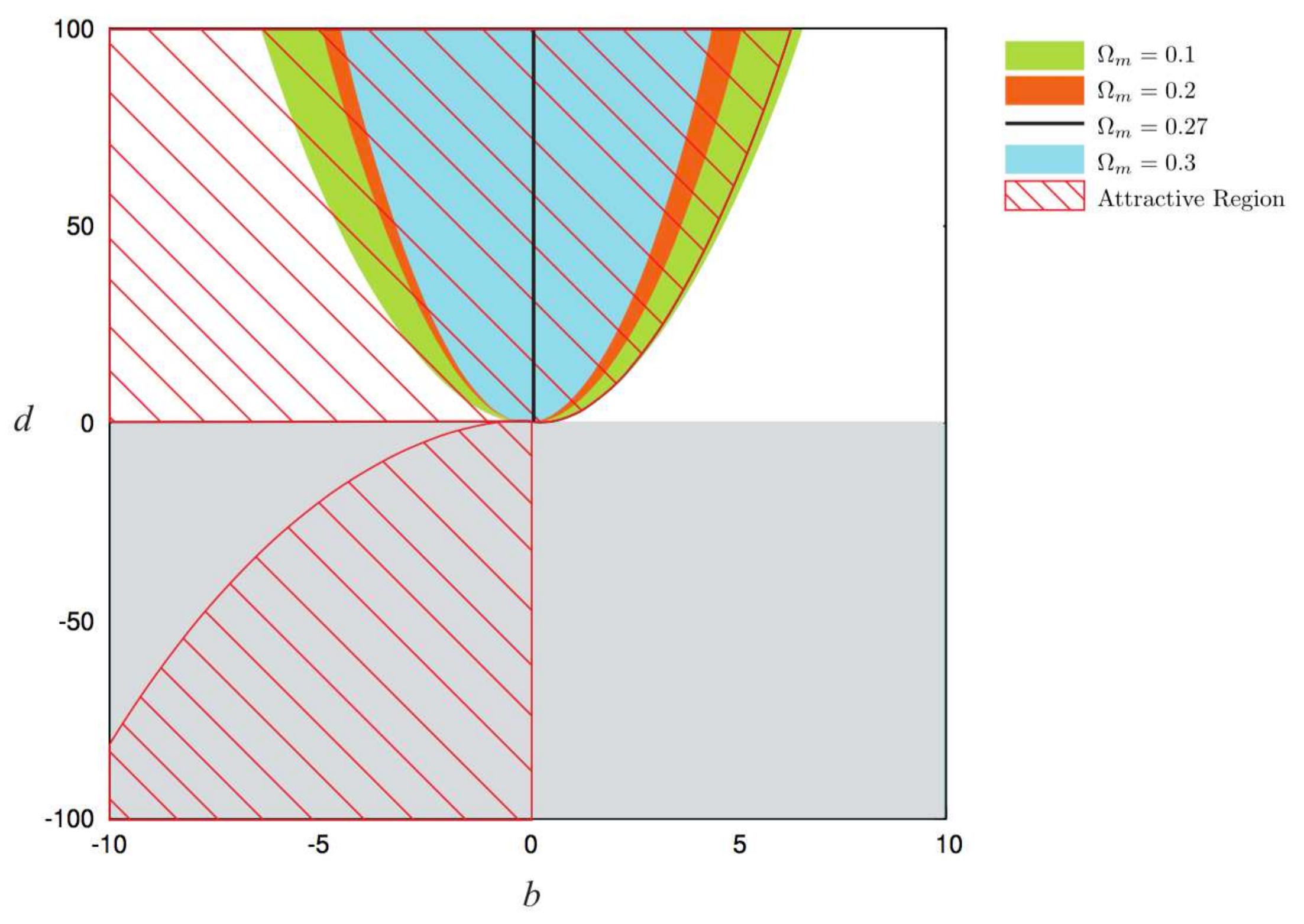} %{n2agsing.eps}
\includegraphics[width=0.47\textwidth]{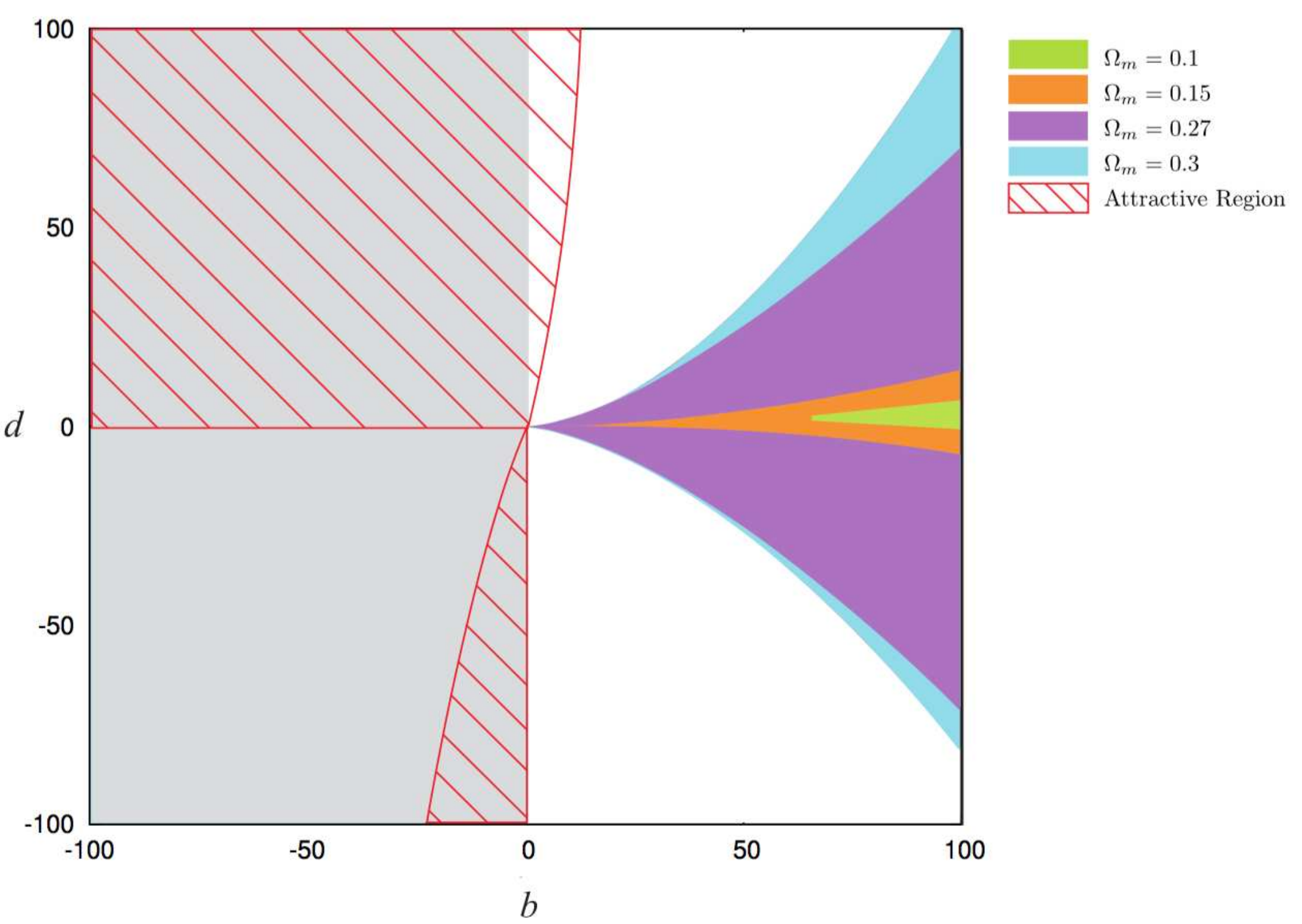} %{n3agsing.eps}
\end{center}
\caption[justification=justified,singlelinecheck=false]{Regions in the $b-d$ plane, for $n=2$ (upper panel) and $n=3$ (lower panel) containing %mostly  
singular/regular sets of parameters for different values of $\Omega_{m}$, and regions with different signs of $R_0$. In both panels, the non-meshed zone represents $R_0>0$, and the grey regions represent entirely singular regions (regardless of the value of $\Omega_{m}$); for $n=2$ this corresponds to $d<0$, and for $n=3$ this corresponds to $b<0$.  Other singular regions in the $b-d$ plane do depend upon the value of $\Omega_m$ and have been represented in different colours (see legends in the panels). Note that this analysis focuses on the past cosmological evolution $z\geq 0$, thus this does not ensure a whole regular condition for $n>1$.
For the case $n=2$, the closer $\Omega_{m}$ gets to its best-fit value $\Omega_m=0.27$ (see Section \ref{Section_SN}), the narrower the aforementioned upper singular parabolic region becomes. When $\Omega_{m}=0.27$, the phase space is completely regular for all values of $d>0$ and all values of $b \neq 0$. For the case $n=3$, there appears not to be any improvement as we get closer to the $\Omega_m$ best-fit value but the singular region located at $b>0$ grows with $\Omega_m$. 

The best fit values found in Table \ref{MCMCtable} lie in the blank regions for both $n=2,3$.}
\label{Fig_3}
\end{figure}

Following the general results in \cite{Dombriz_JCAP_Attractive}, one can prove that 
\begin{eqnarray}
R_{\mu\nu}\xi^\mu\xi^\nu \geq \frac{f(R_0)-R_{0}f_R(R_{0})}{2(1+f_R(R_0))}\;,
\label{Ib}
\end{eqnarray}
where we have just considered Eqn. (\ref{1.2}) with constant scalar curvature and all standard matter sources - if any -  to satisfy the SEC.  Therefore, the r.h.s.\ of \eqref{Ib} must be negative in order to allow $R_{\mu\nu}\xi^\mu\xi^\nu < 0$ or equivalently ${\cal M}_{\xi} > 0$. It follows that ${\cal M}_{\xi}$ must be bounded from above. Hence the necessary condition for timelike geodesics to diverge at late times becomes:
\begin{eqnarray}
\frac{f(R_0)-R_{0}f_R(R_{0})}{2(1+f_R(R_0))} < 0\,,
\end{eqnarray}
and provided that $1+f_R(R_0)>0$, we obtain
\begin{eqnarray}
f(R_0)-R_{0}f_R(R_{0}) < 0\,.
\label{lacondicion}
\end{eqnarray}
If we now consider equation (\ref{1.2}) in vacuum ($T=0$) for constant scalar curvature solutions, the value of $R_0$ satisfies
\begin{eqnarray}
R_0=\frac{-2f(R_0)}{1-f_R(R_0)}\,,
\label{ec_curvatura_constante}
\end{eqnarray}
Although in general this algebraic equation cannot be solved analytically, some $f(R)$ models exist (depending upon the parameters of the model) for which a closed solution can be found. Thus rearranging terms in the equation (\ref{lacondicion}) one can prove that the equation above implies\footnote{Here $1-f_R(R_0)>0$ has been assumed in agreement with usual viability conditions on $f(R)$ theories.}
\begin{eqnarray}
R_0 > 0\,.
\label{Rmayor0}
\end{eqnarray}
Hence, a positive contribution to the Raychaudhuri equation from the space-time geometry ${\cal M}_{\xi}$
for every timelike direction is obtained provided that $R_{0}>0$. This condition will 
constrain the parameters of different Hu-Sawicki models. As mentioned above constant curvatures $R_0$ usually cannot be determined analytically\footnote{For $n=1$,  Eqn. (\ref{ec_curvatura_constante}) can be exactly solved \cite{Dombriz_JCAP_Attractive}.}  from (\ref{ec_curvatura_constante}) although numerical solutions do generally exist, as we shall illustrate for $n=2,3$ in the upcoming section.
\vskip0.2cm
In conclusion, the combination of the analyses described in Sections
 \ref{Sub Singularities} and \ref{Sub Attractive character} provide two complementary independent ways of constraining viable $f(R)$ models. We have applied those results to the Hu-Sawicki class of models for different exponents $n=2,3$ and summarise the results in Fig.  \ref{Fig_3}. 
This information can be then used in MCMC analyses in order to avoid regions in the parameter-space which we know possess singular points in their cosmological evolution or do not provide late-time accelerated expansion.

%%%%%%%%%%%%%%%%%%%%%%%%%%%%%%%%%%%%%%%%%%%%%%%%%%%%%%
\section{Fitting the Hu-Sawicki model with Sne I${\rm a}$}
\label{Section_SN}
%%%%%%%%%%%%%%%%%%%%%%%%%%%%%%%%%%%%%%%%%%%%%%%%%%%%%%
We implement a Markov Chain Monte Carlo (MCMC)  routine to estimate the parameters of the Hu-Sawicki model, by fitting to the Union 2.1 supernovae data (see Ref.~\cite{Suzuki:2011hu}), consisting of 557 sources. Using a Metropolis-Hastings algorithm, we sample from a three-dimensional parameter space $\{b,d,\Omega_{m}\}$, while holding $n$ fixed at three integer values of 1,2,3.  However, the occurrence of singularities in a model's resulting expansion history makes any numerical analysis, such as a parameter optimisation routine, more complicated, as the statistics and posterior distributions may be compromised whenever encountering singular evolutions. We therefore attempt to manage this difficulty as follows.
%%%%%%%%%%%%%%%%%%%%%%%%%%%%%%%%%%%%%%%%%%%%%%%%%%%%%%%
\subsection{Numerical detection of singularities}
%%%%%%%%%%%%%%%%%%%%%%%%%%%%%%%%%%%%%%%%%%%%%%%%%%%%%%%
Given that singularities are expected within the parameter space, it is useful to determine the regions of parameter-space containing regular solutions, so that appropriate priors on the free parameters may be considered.
%or at least, where the presence of singular set of parameters barely appears when studying the cosmological evolution in that phase space.
%
%To be clear, the singularities occur in the first derivative of the Hubble parameter, $h(z)$, and is reflected in the deceleration parameter, $q(z)$. To isolate a ``singular" set of parameters, once the system has been solved, the solution is scanned for any unexpected behaviour, such as the Hubble parameter becoming too small, such that it's reciprocal would be numerically undefined, the deceleration parameter switching sign or becoming too large etc. 

Since the integration of the cosmological equations is needed for the optimisation routine, the detection and avoidance of any singularities is a mandatory step at this stage. %\color{red}{should we not say how we avoid the singularities in the MCMC?}

The case $n=1$ is particularly simple as the appearance of singularities solely depends upon two free parameters $\{b,\Omega_m\}$, and the scalar potential is obtained exactly in (\ref{1.11}). Consequently the region of parameter-space leading to regular solutions can easily be found \cite{SaezGomez:2012ek}. As pointed out above, we need to stay initially on the lower branch of the scalar potential Fig.~\ref{fig2} in order to avoid the singularity, which is located at $\phi=1$ where $V(\phi=1)=\frac{bc}{d}H^2_0$. This leads to the condition:
\be
V_{\phi=1}<\frac{bc}{d}H^2_0\ ,
\label{Sing1}
\ee
Then, by imposing the initial conditions to match the model with $\Lambda$CDM at a particular redshift and using the expression (\ref{1.10}), we get the following constraint on $b$,
\be
b<\frac{3(\Omega_m-1)H_0^2}{R_{0,\Lambda {\rm CDM}}}=\frac{\Omega_m-1}{\left[z_0 \left(z_0^2+3z_0+3\right)-3\right]\Omega_m +4}\ ,
\label{Sing2}
\ee
where $z_0$ is the initial redshift.

For $n>1$, we need to resort to numerical techniques in order to determine the singularity free regions in the parameter space. We proceed by testing a reasonably large grid of the sampling region, within a chosen redshift range. The dynamical system equations (\ref{dhdz}) - (\ref{dodz}) are integrated from $z=10$ to the present era\footnote{As in \cite{DS_Hu-Sawicki}, we see that choosing $z0\geq 10$ provides good initial conditions, since for earlier times, the $\Lambda$CDM model is indistinguishable from the Hu-Sawicki model.%, and it is the late-time dynamical dark energy behaviour which we are interested in exploring with this analysis.
}. 

The solution at every point in the grid is examined to determine any singular behaviour.  We present two-dimensional representations of this grid, showing the $b-d$ plane, for fixed values of $\Omega_{m}$ in Fig.~\ref{Fig_3} for $n=2$ and $n=3$. According to this, for higher values of $n$ the singular regions in the phase space are more complicated. While the filled regions in the plot represent regions which contain the singularities in this range of parameter space, there may exist regular parameter sets for different values of $\Omega_m$ within those regions as well. Similarly, the white space gives singularity free regions that also depend on the value of $\Omega_m$. This analysis ensures an smooth cosmological evolution for $z\geq 0$ but is unable to ensure a future cosmological evolution in absence of singularities. However, note that many other dark energy models allowed by the observations contain future cosmological singularities \cite{SingOc}. 
\subsection{SNIa fit to the Hu-Sawicki model}
%%%%%%%%%%%%%%%%%%%%%%%%%%%%%%%%%%%%%%%%%%%%%%%%%%%%%%
In this section 
we present the results of a Markov Chain Monte Carlo method performed to fit for the free parameters of the Hu-Sawicki model subject to the theoretical constraints presented above. The results obtained in the previous section aid in the avoidance of highly dense singular regions, as well as the interpretation of the MCMC chains. The regions for which the cosmological evolution does not guarantee expansion
are %gravitational interaction becomes non-attractive are 
excluded {\it a priori} in the calculations here, although once the maximum likelihood is obtained, we are able to determine whether the corresponding points in the parameter-phase space lie in the allowed regions, i.e., singularity-free and late-time expansion ones. %{\color{red}Here we have implemented an MCMC by following the Metropolis-Hastings algorithm. For doing so, we have used two independent codes, one using Mathematica and the other one by Python. Each chain has explored $2.5\times10^5$ points of the parameter space. About $20$ chains for each value of $n$ have been worked out in order to analyze the convergence of the chains. The results are described below.}

The observable to be compared with the catalog of Union2 is the apparent magnitude, which is defined as follows
\begin{eqnarray}
m^{th}(z;\Omega_m^0,z_0,x_{i})&=&{\bar M} (M,H_0) +\nonumber\\
&& 5\,{\rm log}_{10} \left[D_L (z;\Omega_m^0,z_0,x_{i})\right]
\label{SN2} 
\end{eqnarray}
where $x_{i}$ are the free parameters of the model and ${\bar M}$ is the magnitude zero point offset, which is given by 
\begin{eqnarray}
{\bar M} = M + 5\, {\rm log}_{10}\left[\frac{c\,H_0^{-1}}{\rm Mpc}\right] + 25\ . 
\label{SN3} 
\end{eqnarray}
Here $M$ is the absolute magnitude and $H_0$ is the Hubble parameter evaluated today, while $D_L (z;\Omega_m^0, z_0 ,x_i)$ is the corresponding free luminosity distance:
\begin{equation}
D_L^{} (z;\Omega_m^0,z_0,x_{i})= (1+z) \int_0^z {\rm d}z'\frac{H_0}{H(z';\Omega_m^0,z_0,x_{i})}\ . 
\label{SN1} 
\end{equation}
Then, for a particular set of the free parameters $\{\Omega_m^0,x_{i}\}$, the Hubble parameter 
$H(z;\Omega_m^0,z_0,x_{i})$ is obtained by solving equations (\ref{dhdz}) - (\ref{dodz}). Thus, the theoretical value of the apparent magnitude (\ref{SN2}) can be determined, and compared with the observational data from \cite{Suzuki:2011hu}, which provides the observed apparent magnitudes $m^{obs}(z)$ of the SN Ia with the corresponding redshifts $z$ and errors $\sigma_{m(z)}$. Then, the best fit is determined by studying the probability distribution
\begin{equation}
P({\bar M}, \Omega_m^0, w_0,z_0)= {\cal N} {\rm e}^{- \chi^2/2}\ , 
\label{SN4} 
\end{equation} 
where  $\chi^2\equiv\chi^2({\bar M}, \Omega_m^0, z_0, x_i)$ and 
\begin{eqnarray}
\chi^2 =\sum_{i=1}^{557} \frac{(m^{obs}(z_i) - m^{th}(z_i;{\bar M}, \Omega_m^0, z_0, x_i))^2} {\sigma_{m^{obs}(z_i)}^2}\,.
\label{SN5} 
\end{eqnarray}
Here ${\cal N}$ is a normalisation factor. Those free parameters $\{{\bar \Omega}_m^0, {\bar z}_0, {\bar x}_i\}$ minimising the $\chi^2$ expression (\ref{SN5}) will correspond to what we call the {\it best 
fit}.  On the other hand, the parameter $\bar{M}$ can be minimised and dropped out of the $\chi^2$ expression. Details on such a process are provided in Appendix \ref{Appendix_I}. 
\begin{table}[htbp]
\begin{tabular}{l | c  |c|| c| c }
\hline
% $n$ & \multicolumn{2}{c}{$b$} & \multicolumn{2}{c}{$d$}  \\ \hline\hline
 $n$ & $b$ & $b_{\rm best\, fit}$ & $d$ & $d_{\rm best\, fit}$ %\multicolumn{2}{c}{$b$} & \multicolumn{2}{c}{$d$}  
 \\ \hline\hline
1 &$347\pm{300}$&  745   &  - &  -  \\ 
2 &$825\pm 200 $ &  1052 &$303\pm{200}$ & 49 \\
3 & $947\pm 300$&1388&$3515\pm 500 $ & 3675 \\ \hline
\end{tabular}
\vskip0.3cm
\begin{tabular}{l | c |c|| c| c }
\hline
% $n$ & \multicolumn{2}{c}{$\Omega_{m}$}  & $\chi_{min}^{2}$ &$\chi_{red}^{2}$ \\ \hline\hline
$n$ & $\Omega_{m}$  & $\Omega_{m\,{\rm best\,fit}}$ & $\chi_{min}^{2}$ & $\chi_{red}^{2}$ \\ \hline\hline
1 & $0.270 \pm 0.020$ & 0.270 & 542.683& 0.979\\ 
2 & $0.272\pm 0.020 $ & 0.270 &542.683 & 0.981\\
3 & $0.264\pm 0.018$ &0.270 & 542.689 & 0.981\\ \hline
$\Lambda$CDM & $0.27 \pm 0.02$ & 0.27 &542.685 &0.978  \\
\hline
\end{tabular}
\caption{MCMC analysis results for the fitting of Hu-Sawicki model to Union 2 SNIa data. The free parameters $b$,$d$ and $\Omega_{m}$ are estimated, for each case where $n$ is fixed, $n=1,2,3$. We include the results for $\Lambda$CDM for comparison. Each free parameter is represented by two columns, the left showing the mean and $1\sigma$ of the resulting posterior, and the right showing its best fit value. 
The best fit values lie in the white regions in Fig. \ref{Fig_3} for both $n=2,3$ exponents, therefore providing the appropriate cosmological expansion behaviour at late time.  }
\label{MCMCtable}
\end{table}
The MCMC analysis for the Hu-Sawicki model was performed by fixing integer values of $n=1,2,3$, sampling for the posterior distributions of the remaining free parameters $b$, $d$ and $\Omega_{m}$. 

For each of $n=1,2,3$, twenty chains were generated, comprising $2.5\times10^5$ sampled points in the respective parameter space. The obvious prior %on $\Omega_{m}$, namely 
$\Omega_{m} \in (0,1]$ was imposed. For the sake of simplicity, each parameter was sampled following a normal distribution centered at zero with standard deviations\footnote{Having initially no information about the scales of $b$ and $d$, the sampling distributions were chosen so as to scan the available phase space efficiently. 
The relatively large values of $\sigma_{b}$ and $\sigma_{d}$ were settled upon in order %ultimately 
to optimise  the computing time. %it is important to note that 
More conservative values for these quantities were tested and the results %- although significantly scaled down - 
did not significantly differ from those presented here.}

\begin{figure*} %[h!]
 \begin{center}
 \includegraphics[width=0.32\textwidth]{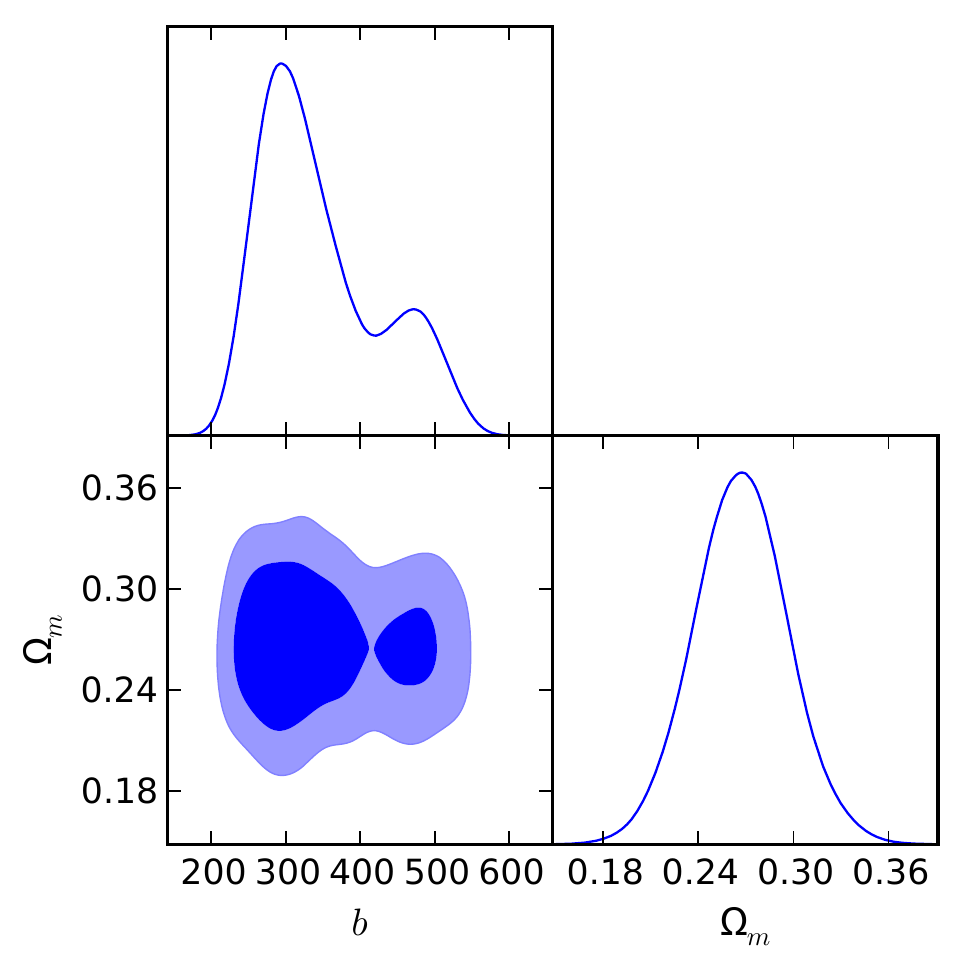}
 \includegraphics[width=0.32\textwidth]{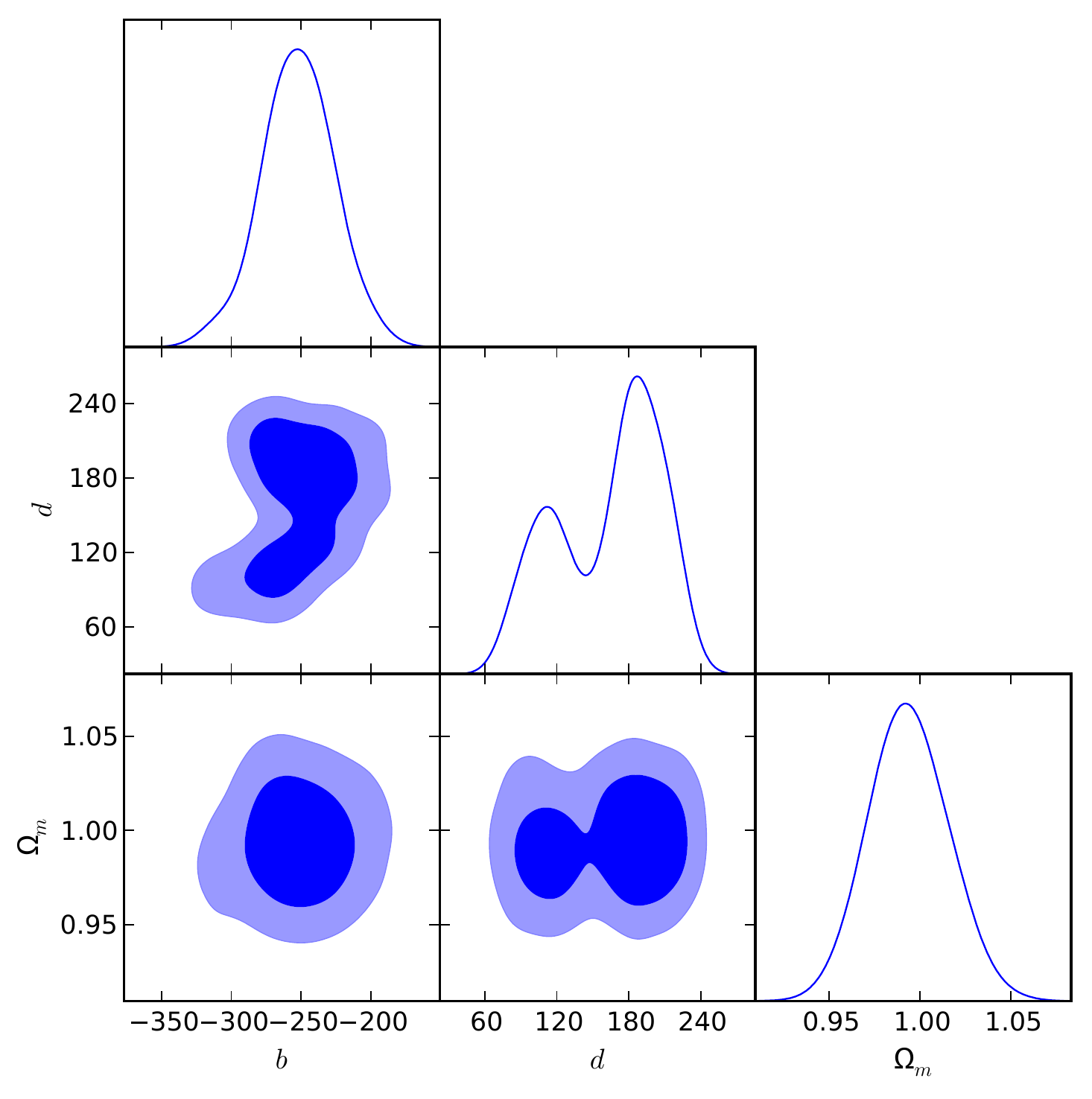} 
  \includegraphics[width=0.32\textwidth]{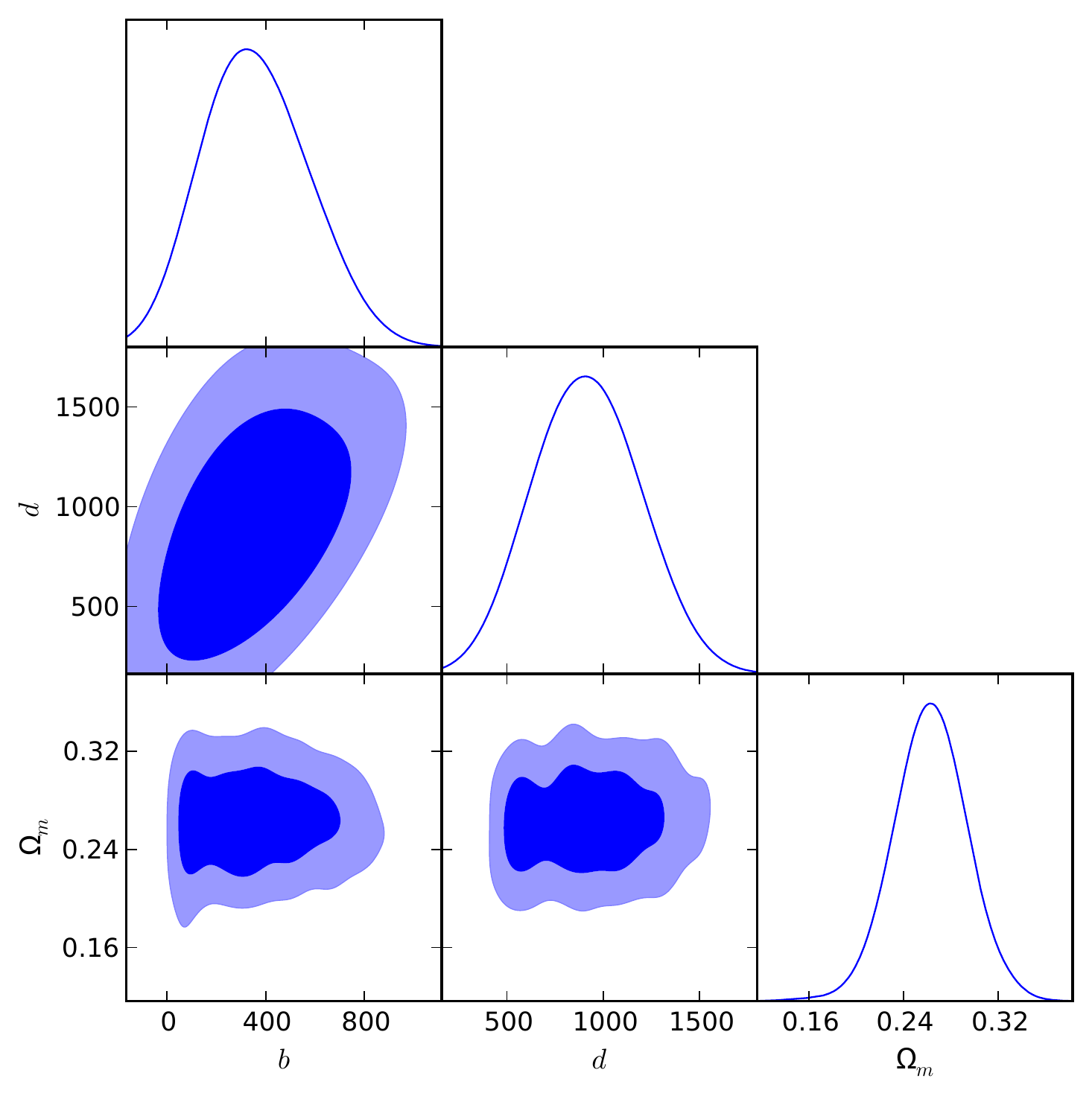}
  \end{center}
  \caption{
 Constraints for the posterior distributions and contours for each free parameter, in each case of $n$, $n=1$ (left panel), $n=2$ (central panel) and $n=3$ (right panel). 
 For $n=1$, parameter $d$ vanishes from the system. We find $\Omega_{m}=0.27 $, and $b$ is unconstrained for a wide range of values. For the case $n=2$, again $\Omega_{m}=0.27$ and $b$ is unconstrained for a wide range of values. %while $d>0$ is also mostly viable. 
For $n=3$, $\Omega_{m}=0.264$, whereas $b$ and $d$ remain unconstrained.}
 \label{triangles} 
 \end{figure*}

$\sigma_{b}=5$, $\sigma_{d}=5$ and $\sigma_{\Omega_{m}}=0.03$ respectively.
%
%The only apparently important constraint on $b$ and $d$, for now, results from seemingly contained regions of the phase space which are over dense with singular, and thus, unviable models. It was important to avoid these regions as far as possible. 
Results are depicted in Fig. \ref{triangles}.
Each chain was initialised at unique points in the phase space, and 
for each Markov Chain, convergence of the matter density fraction of the universe today, i.e., $\Omega_{m}$ occurred fairly quickly.  In fact,  $\Omega_{m}$ is very well described by a Gaussian posterior distribution of all three values of $n$, with an error comparable to that of a similar analysis done for $\Lambda $CDM. Table \ref{MCMCtable} summarises the results for each value of $n$. We include the best fit values for each parameter corresponding to each value of the exponent $n$, as well as the mean and 1-$\sigma$ standard deviation of their sample distribution. We find in all cases that the best fit values do in fact lie in the $R>0$ regions. 

For the case $n=1$, the parameter space is 2-dimensional as $d$ factors out of the system entirely. In this simple scenario, the convergence of the $b$ parameter is remarkably bad (left panel in Fig. \ref{triangles}). We find, consistently for each Markov chain generated, that a range of 
$b$ values minimising $\chi^{2}$ exists. The $\chi^{2}$ surface is extremely flat, and we find that the variation in the  values of the $\chi^{2}$ is small ($\sigma_{\chi^{2}} = 1.470$). 

When $n=2$, the parameter space is 3-dimensional. Once again, $\Omega_{m}$ converges quickly to $\Omega_{m}=0.27 \pm 0.020$, however, $b$ and $d$ show no acceptable convergence in general (mid panel Fig.  \ref{triangles}). In both cases the standard deviations of the posterior distributions are very large. As can be seen from Table \ref{MCMCtable}, the best fit values of $b$ and $d$ are not similar to their mean values. The variation in the $\chi^{2}$ values, $\sigma_{\chi^2}=1.530$, is small in this case as well, implying that a wide range of values for $b$ and $d$ perform similarly when fitting the supernovae data. It is therefore possible for the best fit value, which minimises the $\chi^{2}$, to be quite different to the mean of the posterior. 

Finally, for $n=3$, where $\sigma_{\chi^{2}}=1.364$, it can be seen that the results are very similar to those of $n=2$. Whereas $\Omega_{m}$ successfully converges, $b$ and $d$ remain unconstrained (right panel Fig.  \ref{triangles}). The standard deviations of these two free parameters are large, so that the values which minimise $\chi^{2}$ is not reflected in the statistics of the posteriors. 

At this stage we must emphasise that although all the generated MCMC chains % in this analysis 
gave identical results for $\Omega_{m}$, they provided inconsistent results for $b$ and $d$. The distributions of $b$ and $d$ were highly sensitive to the initial points of the various chains, which reiterates the fact that a wide range of values form part of an acceptable optimum region for the values of $b$ and $d$, some of which are not necessarily connected within the phase space.  We have depicted the chain-dependence of the results for the $\{b, d\}$ parameters in Fig. \ref{all}  showing the results of four different chains for the cases $n=2$ and $n=3$.  As can be seen, $b$ and $d$ show no tendency to converge to a preferred state. We are led to conclude that supernovae data does not impose strong enough constraints on the free parameters of the Hu-Sawicki model.  

%%%%%%%%%%%%%%%%%%%%%%%%%%%%%%%%%%%%%%%%%%%%%%%%%%%%%%
\section{Conclusions}
\label{Section_Conclusions}
%%%%%%%%%%%%%%%%%%%%%%%%%%%%%%%%%%%%%%%%%%%%%%%%%%%%%%
In this paper, we have investigated, through a combination of theoretical and statistical tests, several issues which must be considered when trying to constrain the parameter space of viable $f(R)$ theories of gravity, and by extension any %cosmological 
extended theory of gravity. We focused our study on the Hu-Sawicki model, which is considered to provide a reasonable parameterisation of the required features of effective extensions of the Hilbert-Einstein gravitational Lagrangian which are consistent with the $\Lambda$CDM expansion history and astrophysical tests of gravity. 

We first considered two theoretical constraints, which have been widely overlooked in previous literature, namely the appearance of singularities and upper bounds ensuring the cosmological expansion at late times. As discussed previously, these kinds of viable $f(R)$ models analysed in this manuscript contain a cosmological singularity, where the first derivative of the Hubble parameter diverges \cite{Appleby:2009uf}. By analysing the phase space of this model, particularly in the scalar-tensor framework, we found that the singularity is actually an asymptotically stable point, which can be avoided by an appropriate choice of the free parameters together with convenient initial conditions \cite{SaezGomez:2012ek}. 
%Also we studied the upper bounds for these models to ensure late-time expansion without invoking any energy condition to be imposed on geometrical fluids emerging in extended theories of gravity, but rather relying on the positiveness  of the mean space-time curvature which naturally appears in the Raychaudhuri equation. 
%
%
We then investigated the requirements needed to obtain a positive contribution in the space-time geometry term appearing in the Raychaudhuri equation for time-like geodesics. This upper bound for $f(R)$ models guarantees the non-attractive character of gravity at late-times on cosmological scales, i.e., the cosmological expansion by purely gravitational means. We paid special attention to the asymptotic case of (de Sitter) constant scalar curvature with the sole assumption being that the usual energy conditions for standard fluids hold. The Hu-Sawicki model proved to have free parameters capable of satisfying both constraints. For example, for the exponent $n=2$, when these two analysis were combined, we were able to exclude models with $d<0$ and large regions of parameter space with $b<0$. The necessary conditions for the free parameters which give rise to both singular-free and accelerated de Sitter regimes were presented in Fig. \ref{Fig_3}, where the singular regions are presented for several values of $\Omega_{m}$. For the case $n=3$, we found that we needed to exclude regions where $b<0$ and large regions where $d>0$. In this case, the larger the value of $\Omega_m$, the larger the singular region turns out to be. % and consequently the larger the parameter space needed to be excluded.

\begin{figure*}%[h!]
 \begin{center}
 \includegraphics[width=0.485\textwidth]{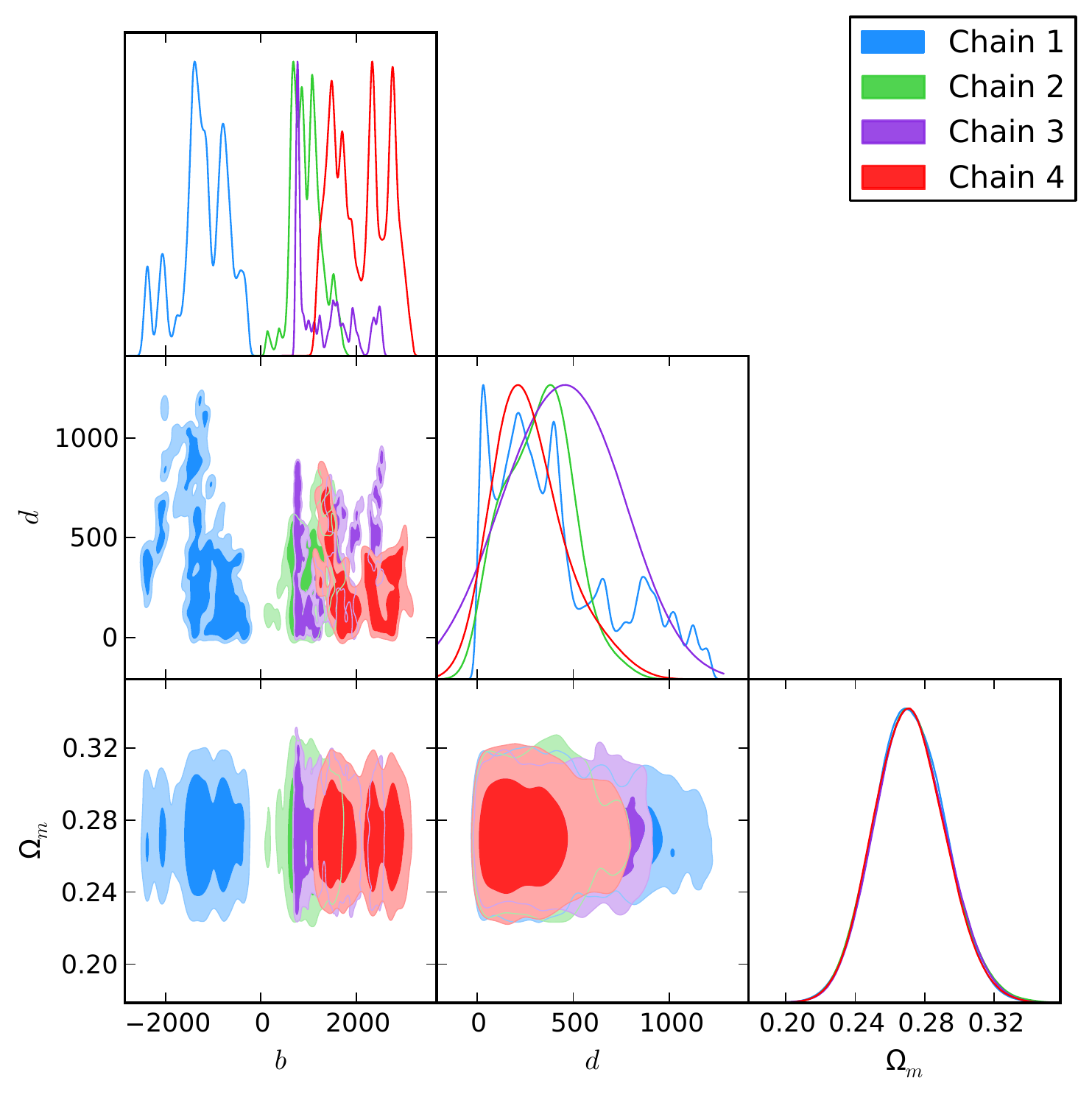}
 \includegraphics[width=0.485\textwidth]{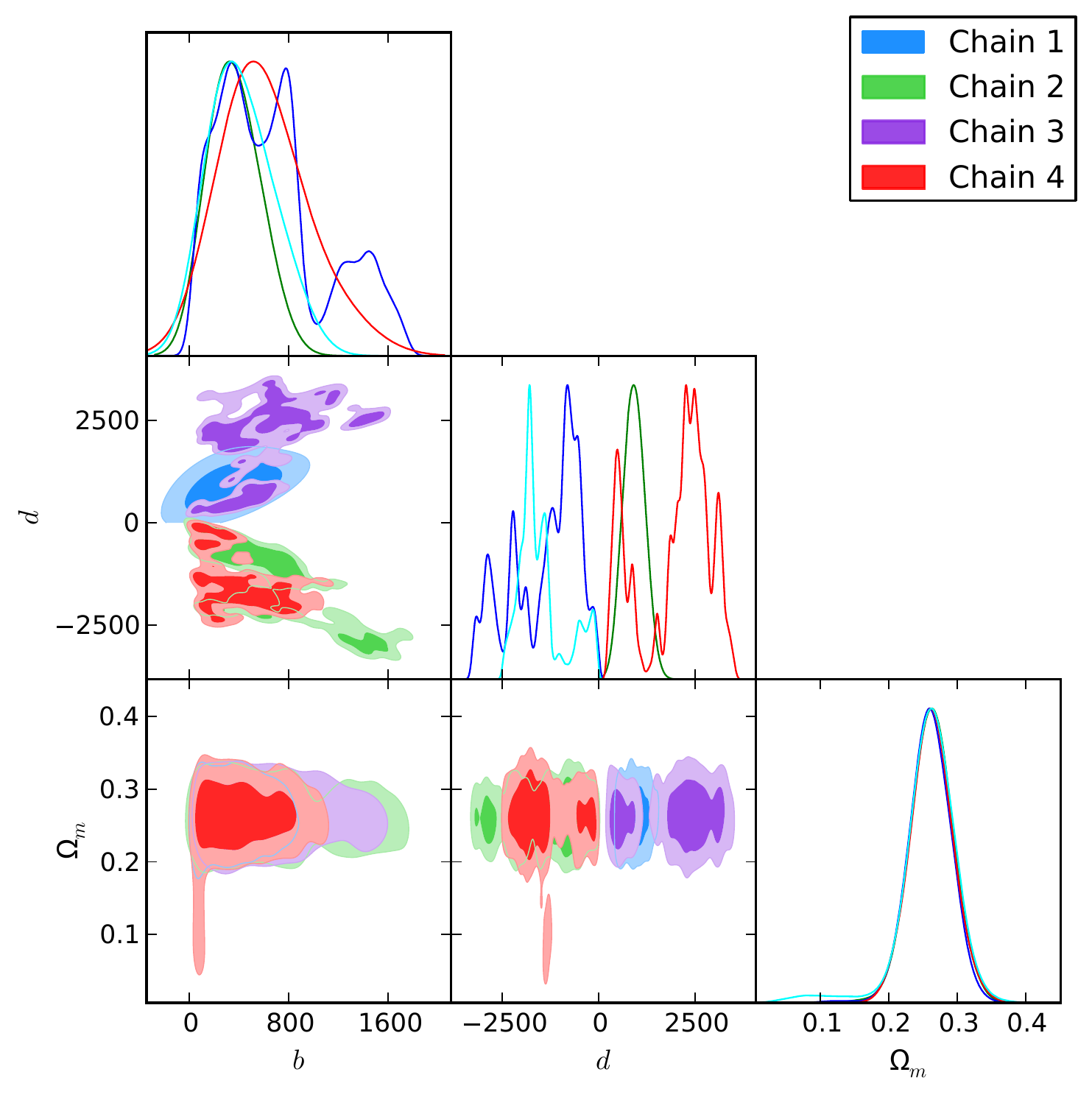}
 \end{center}
 \caption{Above, we plot the results of four MCMC chains corresponding to the cases $n=2$ and $n=3$ to illustrate the lack of hard constraints on the best fit values for the parameters $b$ and $d$ in both cases. }
  \label{all}
% \end{center}
 \end{figure*}

Our aim in this paper was therefore to  constrain the parameters space region that leads to both a smooth and regular Hubble evolution and late-time expansion, and then use these theoretical constraints to determine priors for the free parameters when fitting with Supernovae Ia data.  Using this reduced parameter space, we then looked at what further constraints would be obtained when the expansion history of these models was compared to Supernovae Ia data, using an extensive Markov Chain Monte Carlo analysis. In order to do so, a full resolution of the cosmological background equations was performed using the dynamical system approach \cite{DS_Hu-Sawicki}.
Thus, for exponents $n=2, 3$ the best-fit values that were found for the free parameters lie in both the singularity-free and accelerated regions. 
We also found that while the density parameter of matter $\Omega_{m}$ is well described by a Gaussian posterior distribution of the studied values of exponent $n$, the remaining free parameters $b$ and $d$ cannot be properly constrained by the sole use of supernovae data, with large intervals in the parameter space providing almost the same statistical significance. Consequently, for the studied exponents ($n=1,2,3$), we were not able to improve on what is obtained by the $\Lambda$CDM model. In other words, the supernovae analysis by itself remains a weak tool to constrain $f(R)$ models able to provide an explanation for the accelerated universe beyond the Concordance $\Lambda$CDM model. In fact, this weakness was illustrated by the fact that the best-fit statistical distributions for free parameters did depend upon the starting point of the various Monte Carlo chains as illustrated in Fig. \ref{all}, which gives the results of four different chains for the cases $n=2$ and $n=3$.

We are therefore led to the conclusion that while the theoretical analysis conducted  (the avoidance of  singularities in the cosmological expansion history and the non-attractive character at late times) can indeed be used as a powerful tool to constrain the parameter space of viable $f(R)$ models, when combined with observational constraints coming from supernovae catalogues, does not lead to a significant reduction in the parameter space consistent with a $\Lambda$CDM expansion history. Further observational data, for example large-scale structure surveys, the density contrast at different redshifts and the integrated Sachs-Wolfe effect will be needed in order to improve the exclusion maps we provided in the investigation. Work in this direction is currently in progress.

%Extensions of the analysis presented here can be easily applied to other $potentially viable models in scalar-tensor theories provided a dynamical system %resolution is at disposal to solve the cosmological background evolution %avoiding spurious numerical imprecisions which may mask the real origin of %encountered singularities and bias the extracted conclusions. We referred to %Ref. \cite{DS_Hu-Sawicki} for further details on the Robertson-Walker background %resolution. 
%Theoretical studies in non-homogeneous or anisotropic backgrounds can be also %easily performed \cite{Non-homogeneous}.

%\begin{widetext}
%\end{widetext}

%\begin{widetext}
 %\end{widetext}

%%%%%%%%%%%%%%%%%%%%%%%%
%%%  Acknowledgments
%%%%%%%%%%%%%%%%%%%%%%%%
\section*{Acknowledgments}
We would like to thank David Bacon for a comprehensive reading of the manuscript and for his useful comments.
A.d.l.C.D. acknowledges financial support from the University of Cape Town (UCT) Launching Grants programme and MINECO (Spain) projects 
FIS2014-52837-P, FPA2014-53375-C2-1-P and Consolider-Ingenio MULTIDARK CSD2009-00064.
P. K. S. D. thanks the National Research Foundation (NRF) for financial support.
S.K. is grateful to the NRF and the Faculty of Science, University of Cape Town for financial support.
D.S.-G. acknowledges support from a postdoctoral fellowship Ref.~SFRH/BPD/95939/2013 by Funda\c{c}\~ao para a Ci\^encia e a Tecnologia (FCT, Portugal) and the support through the research grant UID/FIS/04434/2013 (FCT, Portugal). D.S.-G. also acknowledges the NRF for financial support. 

%%%%%%%%%%%

%%%%%%%%%%%

% \bibliographystyle{PRD}

\appendix

%\begin{widetext}
%\section{Coefficients of the differential equation of the evolution of density perturbations}
%\label{APPENDIX A}

\section{}
\label{Appendix_I}

%{\color{red} Here we bla bla bla.} 
In this Appendix we give details on the process to find the apparent magnitude statistical minimum.
Indeed, by expanding (\ref{SN5}) in terms of $\bar{M}$ as 
\begin{equation}  
\chi^2 (\Omega_m^0, z_0, x_i)= A - 2 {\bar M} B  + {\bar M}^2 C\ ,  
\label{SN6} 
\end{equation}
where 
\begin{eqnarray}
A(\Omega_m^0, z_0, x_i)&=&\sum_{i=1}^{557} \frac{(m^{obs}(z_i) - m^{th}(z_i ; \Omega_m^0, z_0, x_i))|_{\bar{M}=0}^2}{\sigma_{m^{obs} (z_i)}^2} \,,
\label{SN7.1} \nonumber \\
B(\Omega_m^0, z_0, x_i)&=&\sum_{i=1}^{557} \frac{(m^{obs}(z_i) - m^{th}(z_i ; \Omega_m^0, z_0, x_i)|_{\bar{M}=0})}{\sigma_{m^{obs}(z_i)}^2} \,,
\label{SN7.2} \nonumber \\
C&=&\sum_{i=1}^{557}\frac{1}{\sigma_{m^{obs}(z_i)}^2 } \,.
\label{SN7.3}
\end{eqnarray} 
The minimum of equation (\ref{SN6}) is located at ${\bar M}={B}/{C}$, such that the $\chi^2$ turns out 
\begin{eqnarray}
{\tilde\chi}^2(\Omega_m^0, z_0, x_i)=A(\Omega_m^0, z_0, x_i)- \frac{B(\Omega_m^0, z_0, x_i)^2}{C} 
\label{SN8}
\end{eqnarray}
Hence, minimising  ${\tilde\chi}^2(\Omega_m^0, z_0, x_i)$ independently of ${\bar M}$, is enough to find the best fit since $\chi_{min}^2={\tilde\chi}_{min}^2$.

\end{document}